\documentclass{aa}

\usepackage{multirow}

\usepackage[usenames]{color}

\begin{document}

\title{Dynamics of the photosphere along the solar cycle from SDO/HMI}

\author{ Th.~Roudier\inst{1}, J.M.~Malherbe\inst{2}, G.M.~ Mirouh\inst{1}}

\date{Received \today  / Submitted }

\offprints{Th. Roudier}

\institute{Institut de Recherche en Astrophysique et Plan\'etologie, Universit\'e de Toulouse, CNRS,
14 avenue Edouard Belin, 31400 Toulouse, France
\and LESIA, Observatoire de Paris, Section de Meudon, 92195 Meudon, France}

\authorrunning{Roudier et al.}
\titlerunning{Dynamics of the photosphere along the solar cycle from SDO/HMI}

\abstract{
  As the global magnetic field of the Sun has an activity cycle, one expects to observe some
variation of the dynamical properties of the flows visible in the photosphere.
}{
  We investigate the flow field during  the solar cycle by analysing SDO/HMI observations
of continuum intensity, Doppler velocity and longitudinal magnetic
field.
}{
    We first picked data at disk center during 6 years along the solar cycle with a
    48-hour time step in order to study the overall evolution of the
    continuum intensity
    and magnetic field. Then we focused on thirty 6-hour sequences of quiet regions without
    any remnant of magnetic activity separated by 6 months, in summer and winter, when disk center latitude
$B_0$ is close to zero. The horizontal velocity was derived from the
local correlation tracking technique over a field of view of 216.4
Mm ~ $\times$ ~ 216.4 Mm located at disk center.

}{
 Our measurements at disk center show the stability of the flow properties between
meso- and supergranular scales along the solar cycle.
}{
 The network magnetic field, produced locally at disk center
independently from large scale dynamo, together with continuum
contrast, vertical and horizontal flows, seem to remain constant
during the solar cycle. }


\maketitle

\section{Introduction}

The Sun is a variable star with an 11-year activity and a 22-year
magnetic field cycle. The activity cycle is mainly due to the
emergence of sunspots through the photosphere, produced by the solar dynamo. 
That layer reflects the dynamical properties of the
upper convection zone where solar granulation and supergranulation
\citep{RR10} are observed. The subtle and complex interactions
between solar convection and magnetic fields are the basic elements
of the solar cycle. These elements must be observed in detail at
different space and time scales, in order to investigate the large
puzzle of various mechanisms which generate the recurrent
activity. Interactions between the magnetic elements and the
turbulent convective motions contribute to the diffusion of the
magnetic field over the solar surface during the cycle
\citep{UMTV2016,Stein2012,MZ2009}. The variation of the properties
of convective structures along the cycle is thus essential to learning 
more about the quiet Sun. Observations of their dynamical behavior
emphasize the relationship with evolutions of the supergranular
network. Previous works have been performed to detect density or
size variations in the granulation, supergranulation and quiet
magnetic network. The chronology of the study of  solar granulation variations
is detailed by \citet{MHS2007}, \citet{MSH2006} and \citet{RR1998}.
These investigations reveal the difficulty in keeping the same
criterion to characterize the variability of the granulation. The
recent improvement of data homogeneity and processing indicate that,
if there is any change in granulation properties, it is smaller than
3\% (Muller, private communication).

 Supergranulation evolution with solar activity has been
studied by various means, such as flow properties (Doppler velocity,
divergence field, helioseismology) or the network magnetic field. The
association between supergranulation and the magnetic network gives
a natural argument to use such a proxy. Flow analyses
\citep{MRR2008,MRT2007,DET04} show that the correlation between the
network size and activity is sensitive to the definition of the
activity level and do not allow conclusions to be made on how the 
supergranulation varies. The network variations along the solar
activity cycle remain difficult to establish observationally, with
many studies yielding contradictory results (for details see
\citealp{TCB2014}). From the helioseismic side, the dispersion relation
for the supergranulation oscillations appears to be only weakly
dependent on the phase of the solar cycle \citep{GD2004}. However
they reported a decrease of the lifetime and power anisotropy of the
pattern from solar minimum to maximum.

 The variation of the surface dynamics could have a potential effect on
the production of the network and internetwork magnetic elements in
the quiet Sun  \citep{GBOKD14}. One consequence could be an effect
on the total irradiance variation along the 11-year solar cycle
\citep{DEJK2016}.  In addition, our analyses provide a potential method
to distinguish between a surface dynamo and the global dynamo field as sources
of magnetic elements seen in the photosphere.

 The SDO/HMI instrument generates continuum intensity, line of sight (LOS)
Doppler velocities (VLOS) and magnetic fields (BLOS) which can
 be extracted from homogeneous and long temporal sequences. Observations allow 
study of the dynamical properties along the solar cycle over a large field of view (FOV).

Section 2 summarizes overall properties of the continuum intensity and
BLOS at disk center using data from 2010 to 2016 with a 48-hour step
and demonstrates the role of disk center latitude ($B_0$) and apparent
radius variations.

Section 3 presents thirty 6-hour sequences at disk center for
$B_0 =0$ recorded at 6-month intervals from 2010 to 2015
together with
data reduction to derive horizontal flows.

Then, horizontal velocities and their spatial and temporal
properties are described in section 4.

In Section 5 we present the evolution of the magnetic field,
intensity contrast and Doppler vertical velocities along the cycle.

We finally discuss in section 6 the stability of the flows we found
along the solar cycle.

\begin{table}
\begin{center}
\begin{tabular}{|ccc|}
\hline
\multirow{2}{*}{2010}&June     &6, 7\\
                     &December &6, 7\\
\hline
\multirow{2}{*}{2011}&June     &8, 9, 10, 11, 12, 13\\
                     &December &11, 16\\
\hline
\multirow{2}{*}{2012}&June     &10, 11\\
                     &December &7, 8\\
\hline
\multirow{2}{*}{2013}&June     &4, 5\\
                     &December &8, 9\\
\hline
\multirow{2}{*}{2014}&June     &1, 2\\
                     &December &4, 5\\
\hline
\multirow{3}{*}{2015}&May      &29, 30\\
                     &October  &15, 16\\
                     &December &5, 6\\
    \hline
    \end{tabular}
  \caption{Observation dates. The angle $B_0$ lies between -1.05$^\circ$ and 0.61$^\circ$ except for October 2015 where $B_0 = 5.9^\circ$.}
  \end{center}
\end{table}

\section{Intensity contrast and magnetic field along the cycle at disk center}

\subsection{Continuum intensity contrast and solar cycle}

\begin{figure*}
\begin{minipage}[c]{.45\linewidth}
\begin{center}
\includegraphics[scale=0.5]{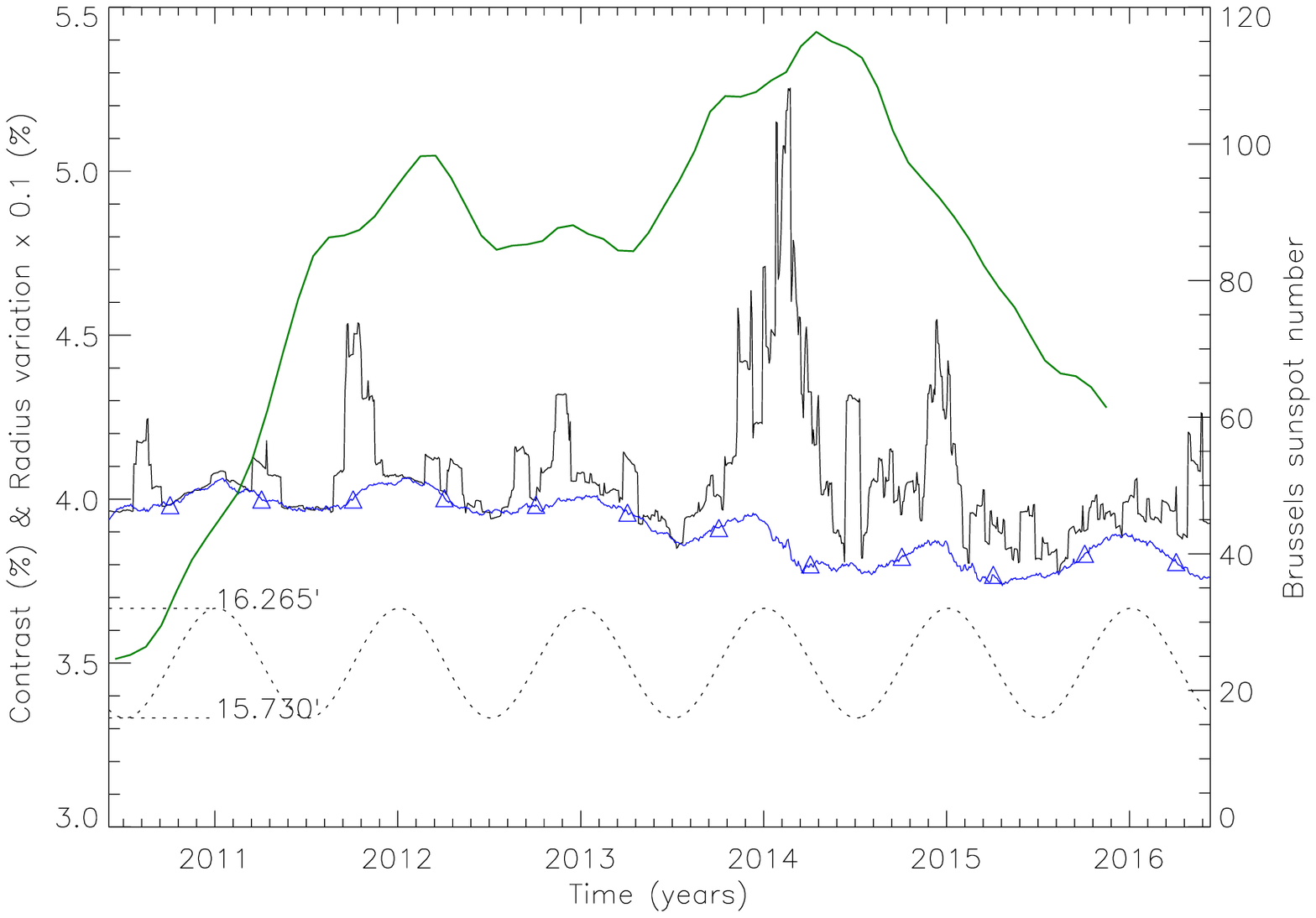}
\caption{black line: contrast of the FOV as a function of time; blue
line: contrast after rejection of structures darker than 90\% of
mean intensity; triangles:  contrast where the apparent radius
(from ephemeris) is $16.0\arcmin$} ; dotted line: solar apparent
radius (arc min); green line: Brussels sunspot index.
\label{icont}
\end{center}
\end{minipage}
\hfill
\begin{minipage}[c]{.45\linewidth}
\begin{center}
\includegraphics[scale=0.5]{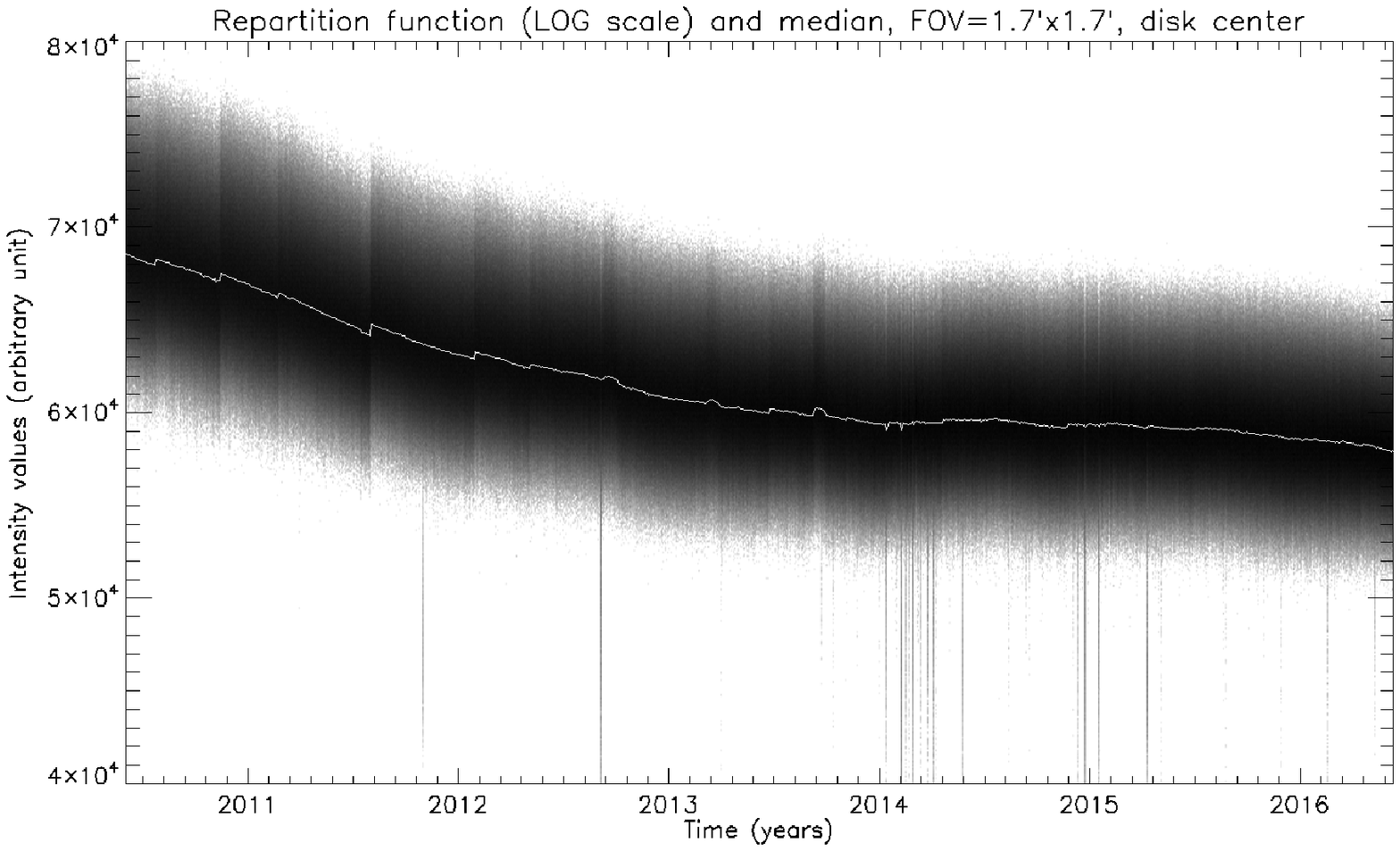}
\caption{ Time evolution of the continuum intensity histograms
and median value.} \label{icont2}
\end{center}
\end{minipage}
\end{figure*}

The Helioseismic and Magnetic Imager \citep{Scherrer2012,Schou2012}
onboard the Solar Dynamics Observatory (SDO/HMI) provides
uninterrupted full disk observations. We first extracted the continuum intensity 
of the 617.3 nm FeI spectral line from HMI data. We
selected a FOV of $5.7\arcmin \times 5.7\arcmin$ at disk center
(pixel size 0.5\arcsec) and selected data along the solar cycle from
June 1, 2010 to June 20, 2016 (6 years) with a time step of 48
hours.

Figure~\ref{icont} shows the monthly averaged contrast of the
intensity continuum of the FOV along the solar cycle. This curve is
well correlated to the Brussels sunspot index, because pores or
sunspots sometimes appear close to the disk center. The contrast is
defined as the ratio of the root mean square (RMS) intensity to the
mean intensity over the FOV. We also show the contrast after
elimination of most of the pores and sunspots: for that purpose,
structures with intensity smaller than 90\% (blue curve) of the mean
FOV intensity were rejected. The contrast is modulated by the
apparent variation of the solar radius (3\% from $15.730\arcmin$ to
$16.265\arcmin$); as a consequence, the pixel size varies, so that
we expect an annual fluctuation of the contrast. The figure effectively 
reveals that the contrast (after pore or sunspot rejection)
presents absolute fluctuations of approximately 0.1\% or relative
fluctuations  3\% in phase with the apparent radius. At this stage,
intensities were not filtered from photospheric 5-min oscillations,
and satellite motion was not corrected either, which can slightly
affect contrasts. Points  where the apparent radius (from
ephemeris) is $16.0\arcmin$ (every 6 months, (triangles)) show that
the granulation contrast is approximately 3.5\% along the solar cycle with
no significant variation.  The measured contrasts are reduced by
a factor of 2.5 due to the instrumental PSF which is found after
restoration up to $9-13\%$ \citep{YEO2014}. Figure~\ref{icont2}
shows the evolution in time of the intensity histograms and indicates 
a slowly decreasing sensitivity of the CCD. This variation does not, however, 
affect the contrast measurements.

\subsection{Line of sight magnetic field and solar cycle}

\begin{figure}
\centering
\includegraphics[width=9cm]{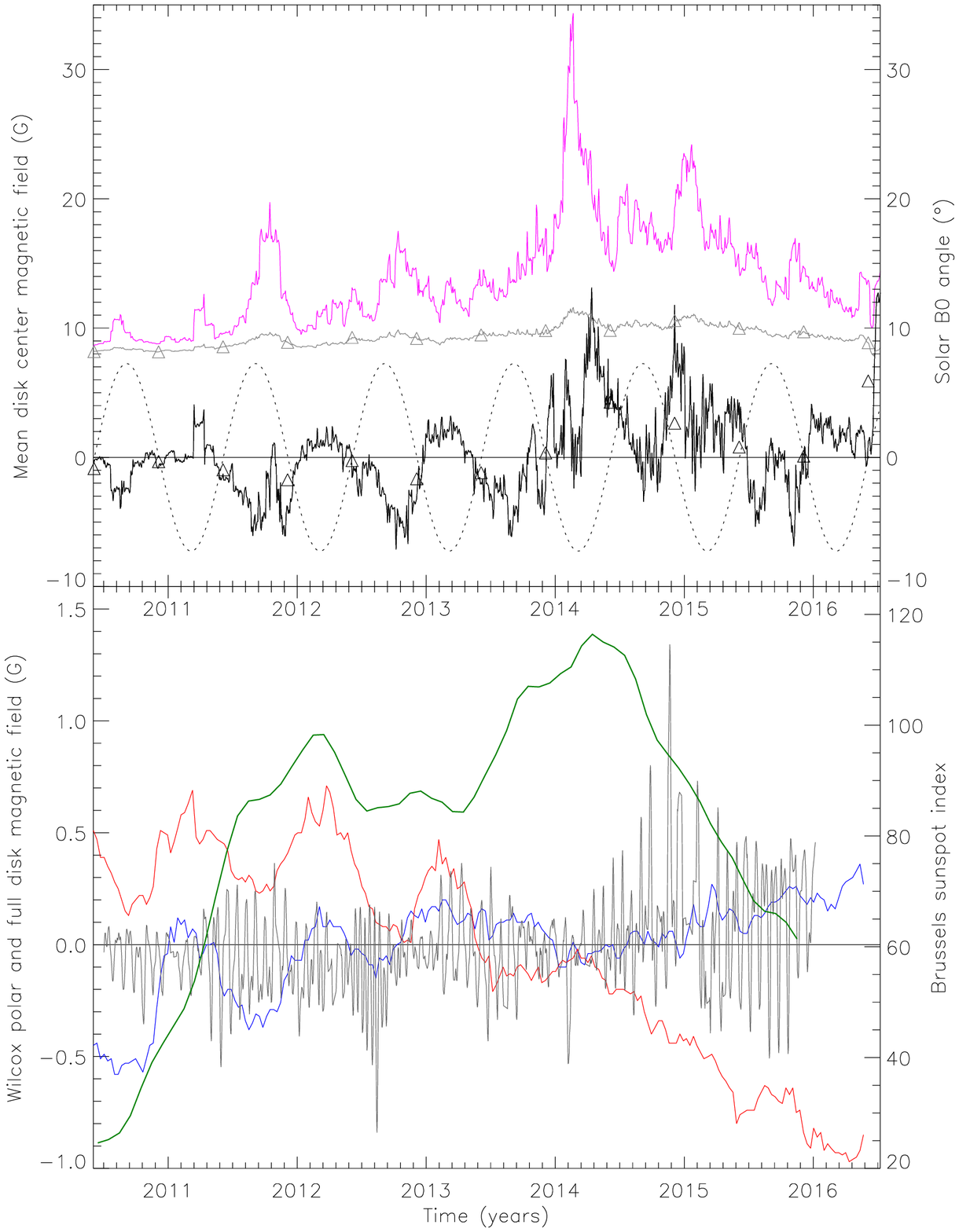}
 \caption[]{  Top panel;
black line: BLOS (Gauss) of the FOV as a function of time;
violet line: mean absolute value of BLOS (Gauss) of the FOV as a
function of time;
gray line: same for regions with |BLOS| < 100G;
triangles: same, but at zero B0 angle;
dotted line: disk center latitude B0 from ephemeris (degrees).

Bottom panel; blue line: North pole BLOS (Gauss) from Wilcox; red
line: South pole BLOS (Gauss) from Wilcox; black line: mean full
disk BLOS (Gauss) from Wilcox; green line: Brussels sunspot index.
} \label{bmagcyc}
\end{figure}

We extracted BLOS measured in FeI 617.3 nm spectral line at disk center, for the
same FOV and times (June 1, 2010 to June 20, 2016, 48-hour time
step), together with continuum intensities from HMI data.Here we used a reduced 
area of $1.7\arcmin$ x $1.7\arcmin$) in order  to limit structures as pores or spots 
as best as possible.

Figure~\ref{bmagcyc}  (top) shows the monthly averaged BLOS
(black) observed in the FOV along the solar cycle and Brussels
sunspot index. We also show the mean BLOS magnitude (violet line) of
the FOV after elimination of active regions (gray line): for that
purpose, regions with $|BLOS| > 100 G $ were rejected. BLOS is
strongly modulated by disk center latitude $B_0$ variations (-7.25
to 7.25 degrees). This effect is due to the annual variability 
of the apparent inclination of the Earth orbit with respect to the 
solar rotation axis. The polar magnetic field from Wilcox at North and
South poles  Figure~\ref{bmagcyc}(bottom) also exhibits the
same $B_0$ dependance, however there is an asymmetric reversal at solar
maximum, as expected, in 2013, lasting one year. Points at zero
$B_0$ angle (every 6 months, triangles, Figure~\ref{bmagcyc}
(top) show that the magnetic field in the quiet network is almost 
constant along the solar cycle with no significant variation ( in
particular regions where $|BLOS| < 100 G $). We therefore worked at
zero $B_0$ in sections 4 and 5 to improve the results.

 \section{Observations at zero $B_0$ angle and data reduction}

The high cadence full disk observations of SDO/HMI give a unique
opportunity for mapping surface flows on various scales (both
spatial and temporal).
 Table 1 summarizes the thirty time sequences we selected from HMI
 continuum intensity, Doppler and BLOS. Particular emphasis was  put 
on only selecting dates where the magnetic field does not contain any
trace of remnant activity. Such regions are supposed to represent
the quietest parts of the Sun.

The selected dates have $B_0$ angle close to zero to avoid
projection effects as emphasized by section 2. The dates run from
2010 to 2015 around June and December of each year.  The $B_0$ angle
values lie between $-1.05\degr$ and $0.61\degr$ except for the 15
and 16 October 2015 where $B_0$ is approximately $5.9\degr$. We used
the 45 sec time series denoted $hmi.Ic\_45s$. This continuum
intensity is provided by the fit to the measurements at six tuning
positions of HMI during the line scan. All the time sequences were 
selected between 0 h and 6 h U.T.. The solar rotation has
been removed along sequences in order to superimpose the data by
applying a temporal shift equivalent to the equatorial rotation of 2
km/s.

The thirty 6-hour time sequences are located  at disk center where
we extracted a final FOV of 216.4 Mm~ $\times$~ 216.4 Mm  for all dates
giving different sizes of field in pixels in the summer or winter
season due to the solar radius variation. The
longitudinal magnetic field and Doppler data were selected only for
one hour at each date. To remove the effects of
the oscillations, we applied a subsonic Fourier filter. This filter
was defined by a cone in the $k$-$\omega$  space, where $k$ and
$\omega$ are spatial and temporal frequencies. All Fourier
components such that $\omega/k\geq V_{\rm
cut-off}=6\,\mathrm{km~s}^{-1}$ were removed to keep only convective
motions \citep{TTTFS89}. We derived horizontal velocity fields from
image granulation tracking using the Local Correlation Tracking
(LCT, \citealp{NS88}).
  Our data processing took into account the 3\% variation of the
pixel size on the Sun surface from 371 km in June to 360 km in
December, due to the different distance between SDO and the Sun, as
noticed in section 2. This correction was directly applied to 
the horizontal velocity amplitudes.

\section{Horizontal motions derived from local correlation tracking at zero $B_0$}

    The horizontal velocity with the LCT technique were computed with a 
time-window of 30 min and spatial scale window of 2.5 Mm ($3.5\arcsec$).  
Each 6-hour sequence corresponds to twelve velocity maps. Horizontal 
velocities computed from the 5-min filtered solar granulation sequence 
have been corrected from SDO proper motions (projected velocities on the 
Sun of OBS\_VR, OBS\_VW and OBS\_VN). As they are also corrected from the 
solar rotation and 5-min oscillations, our measurements reflect the convective 
flow properties.

\begin{figure}
\centering
\includegraphics[width=9cm]{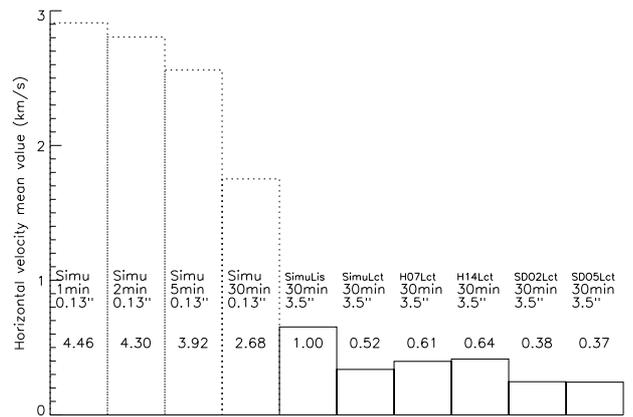}
 \caption[]{ Columns 1 - 4: plasma velocity of simulation for 1, 2, 5, 30 minutes
 and $0.13\arcsec$ time and space resolutions;
column 5: plasma velocity of simulation for 30 minutes and
$3.5\arcsec$ resolutions (LCT-like); column 6: LCT velocity of
simulation; columns 7-8: HINODE LCT velocity for years 2007 and
2014; columns 9-10: SDO LCT velocity, averaged from 2010 to 2015,
for two different FOVs of respectively $2\arcmin$ and $5\arcmin$. }
\label{LCTstat}
\end{figure}

It is well known that the LCT technique underestimates the real
velocities due to data smoothing by a correlation window
\citep{RRMV99,GZKB2007}. In order to calibrate horizontal velocities
obtained from the LCT algorithm, we used the results of a quiet Sun
3D MHD simulation by \citet{Stein2009}. With a $0.065\arcsec$ pixel, their 
simulation provides the three components of the velocity field and magnetic 
field as a function of optical depth and time. We selected the horizontal 
component ($v_x, v_y$) of the velocity at the
solar surface at $\tau=1$  with a $0.13\arcsec$ pixel and one minute time
resolution. Space and time resolutions were both degraded to the LCT
windows of $3.5\arcsec$ (2.5 Mm) and 30 minutes using Gaussian
smoothing. LCT horizontal velocities were computed from intensity
structures (granulation) provided by the simulation. We compared
this result with LCT velocities determined in the quiet Sun at disk
center by HINODE (blue continuum at $0.11\arcsec$ and  at $\tau=1$
and SDO continuum, at $\tau=1$, at $0.50\arcsec$ pixel size, as described above.
 The three data sets are all observed and simulated at the intensity continuum
at $\tau=1$ corresponding to the same height of formation.
Figure~\ref{LCTstat} shows the comparison between plasma velocities
derived from the simulation at the LCT resolution and LCT
velocities issued from structure tracking, for the simulation as
well as HINODE and SDO. We found that LCT results are consistently 
underestimated by a factor 0.52, 0.61 and 0.38  for the
simulation, HINODE and SDO, respectively, so that in the following, the SDO LCT
horizontal velocities were multiplied by the a factor of 2.63.

\begin{figure}
\centering
\includegraphics[width=9cm]{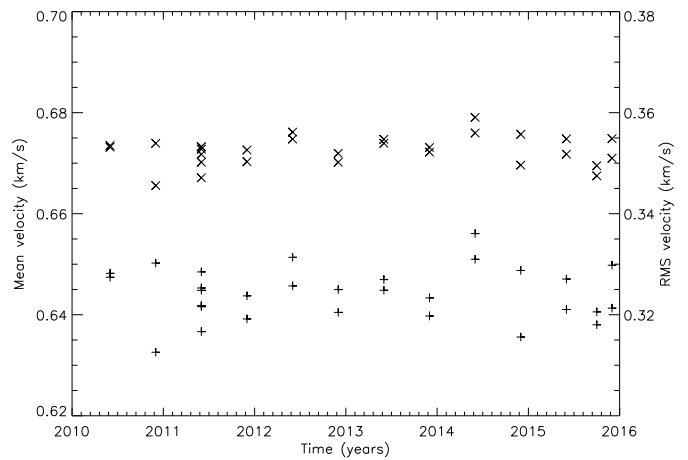}
 \caption[]{Mean magnitude (plus signs) and RMS (crosses) of horizontal velocities from 2010 to 2015.}
\label{Vhmean}
\end{figure}

\begin{figure}
\centering
\includegraphics[width=9cm]{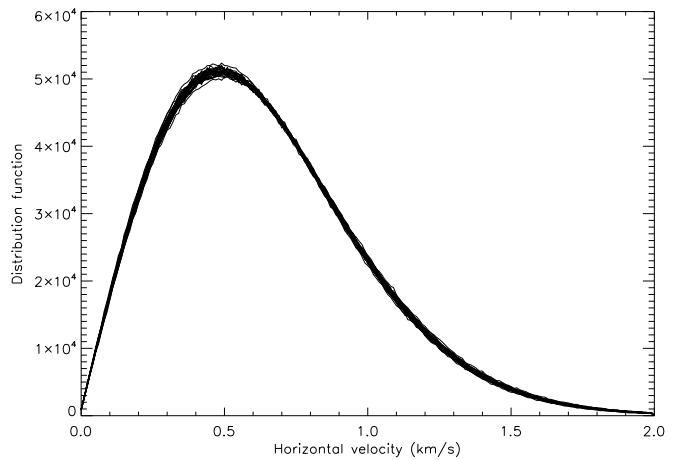}
 \caption[]{Histograms of horizontal velocities of the thirty sequences from 2010 to 2015.}
\label{Vhisto}
\end{figure}

Figure~\ref{Vhmean} shows that during the cycle the mean velocity
magnitude over the 6-hour sequence is $0.645 \pm 0.010$ km/s
while the median value (not displayed) lies at $0.595 \pm 0.010$
km/s. No significant variation is observed during the period 2010 to
2015 although maximum solar activity is reached in 2014. We also found that 
the root mean square (RMS) of the horizontal velocity
remains stable at approximately $0.355 \pm 0.010$ km/s during the solar cycle.
This stability is confirmed by velocity histograms over the cycle that exhibit 
quite similar shapes with very little dispersion (Figure~\ref{Vhisto}).
 The contrast does not reveal any variation in time, corroborating preliminary
conclusions of section 2 and previous results found by \citet{MHS2007}.
(private communication).

\subsection{Divergence and vorticity properties }

\begin{figure}
\centering
\includegraphics[width=9cm]{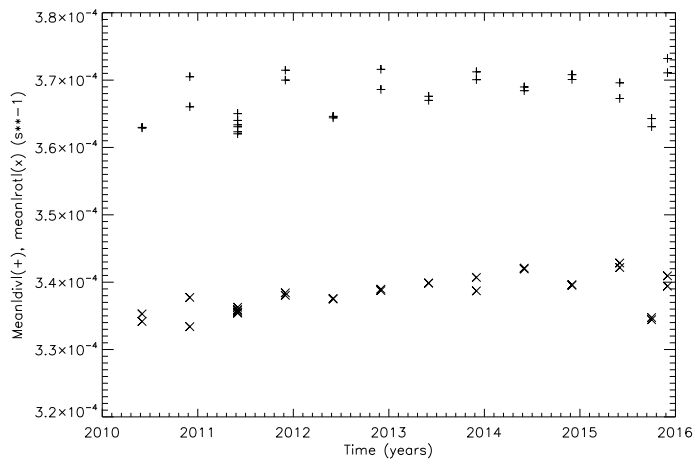}
 \caption[]{Evolution of mean divergence (plus signs) and vorticity (crosses) magnitudes
 of horizontal velocities from 2010 to 2015.} \label{divrot}
\end{figure}

 As the measured velocity field is purely two-dimensional, two quantities are relevant to
characterize flow structures: the divergence
\mbox{$D=\partial_xv_x+\partial_yv_y$} and the $z$-component of the
vorticity \mbox{$\zeta=\partial_xv_y-\partial_yv_x$}.

Figure~\ref{divrot} reveals that both divergence and rotational
magnitudes do not present any significant variation, indicating
again that flow properties remain similar over the FOV during the
solar cycle. The mean values are  $3.68 ~ 10^{-4} ~
s^{-1}$ for the divergence and $3.38 ~ 10^{-4} ~ s^{-1}$ for the
vorticity. We notice a small increase in vorticity between 2010
and 2015,  which does not correlate with solar activity.

\subsection{Diffusion index}

\begin{figure}
\centering
\includegraphics[width=9cm]{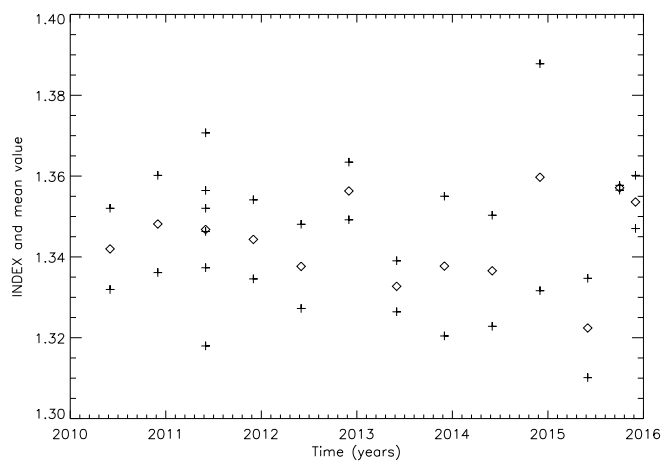}
 \caption[]{Evolution of the diffusion index from 2010 to 2015.
 Mean values for each date are represented by diamonds.}
\label{index}
\end{figure}

\begin{figure}
\centering
\includegraphics[width=9cm]{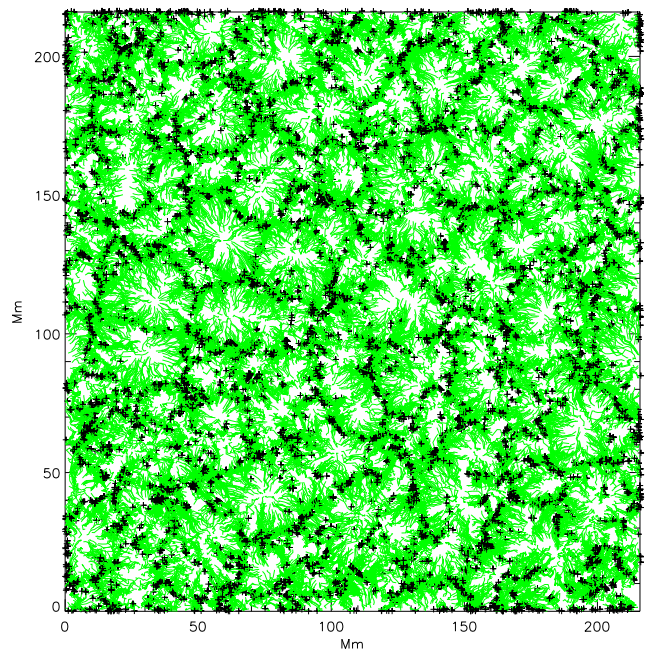}
 \caption[]{Trajectories of corks (green lines) and their final locations (black crosses) after 6-hour
evolution. The ending positions delineate the network.}
\label{corks}
\end{figure}

 The transport of the magnetic element on the photosphere is well
described by a power law $<(\Delta r)^2>=c~t^\gamma$ where $<(\Delta
r)^2>$ is the mean square displacement, $c$ a constant, $t$ the
time measured since the first detection, and $\gamma$ the
spectral index \citep{GSBDB2014}. Following passive scalars, such as 
corks over a long time scale, allows for measurement of this reference
index. Figure~\ref{index} shows that the mean index (diamond) lies
in the range 1.32$-$1.36 but we do not detect any variation
correlated with the cycle, indicating that quasi-stable flows
occur at the Sun surface and drive magnetic elements to form the
quiet network (Figure~\ref{corks}).

\subsection{Cell properties along the cycle}

\begin{figure}
\centering
\includegraphics[width=9cm]{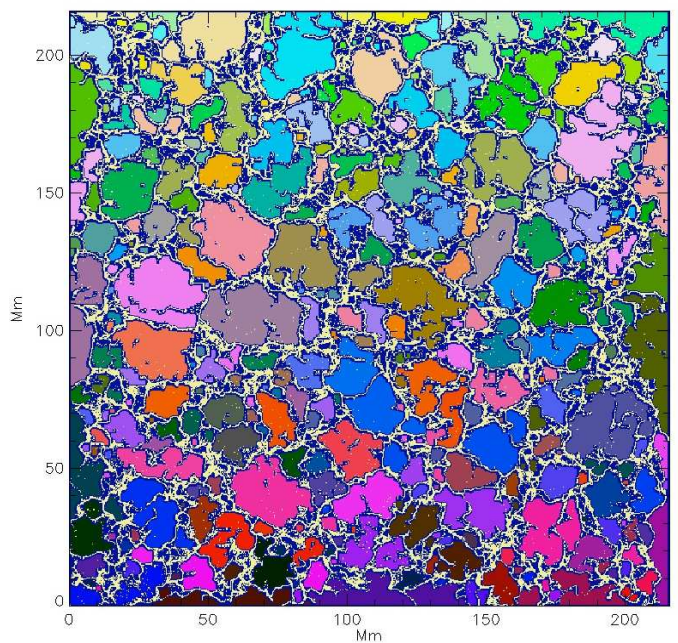}
 \caption[]{ Network location (yellow for cork final positions) surrounding the detected cells shown in
 various
colors.} \label{cells}
\end{figure}

\begin{figure}
\centering
\includegraphics[width=9cm]{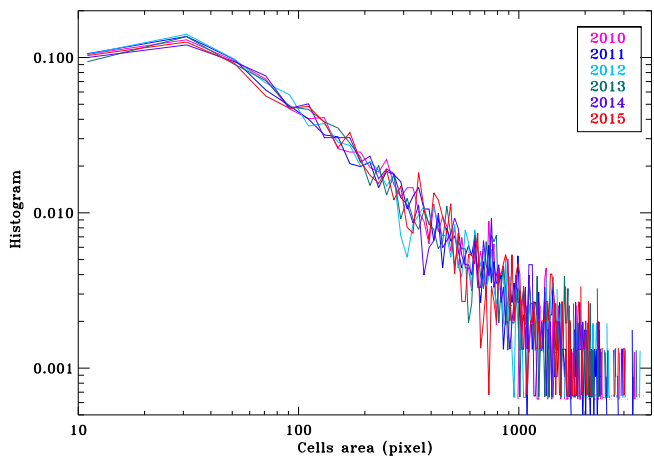}
 \caption[]{Cell area histograms from 2010 to 2015.}
\label{histocells}
\end{figure}

 The cork trajectories due to horizontal velocities have been computed over 
6 hours for each sequence. We observe that the corks are expelled
from diverging cells with size from meso- to supergranular scales as
expected. The corks locations at the end of the sequence
(Figure~\ref{cells}) delineate a network from which we extracted
cells by a segmentation process \citep{RRMV99}.  As the
temporal sequence is relatively short (6 hours), smaller cell sizes are
privileged but the same treatment is applied to all sequences
allowing a quantitative comparison. Figure~\ref{histocells} exhibits
the mean cell area histograms (i.e., the fraction of the number
of cells for each area bin) for each year. We observe a monotonic
decrease towards the larger scales but the different histograms  
are similar, confirming that the size of cells formed by
horizontal flows does not vary significantly over the solar cycle in the
quiet Sun.

\section{Magnetic field, intensity contrast and Doppler properties during the cycle at zero $B_0$}

\subsection{Magnetic field and intensity contrast temporal evolution}

\begin{figure}
\centering
\includegraphics[width=9cm]{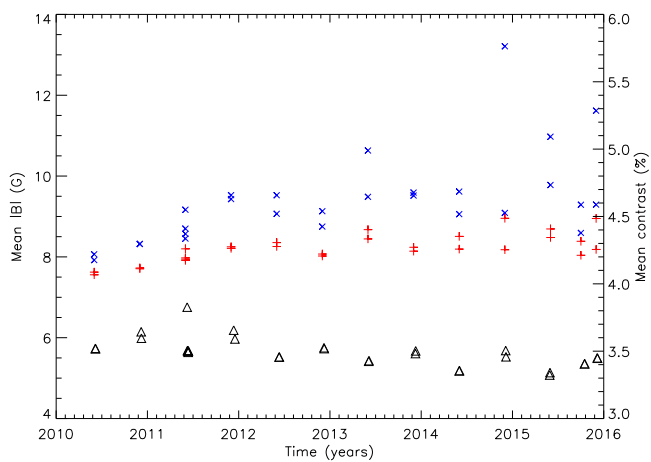}
 \caption[]{Time evolution of the magnitude of BLOS (blue crosses),
 magnetic field in regions where $|BLOS| < 100G$ (red crosses) and intensity contrasts (black triangles).}
\label{Bmoy}
\end{figure}

\begin{figure}
\centering
\includegraphics[width=9cm]{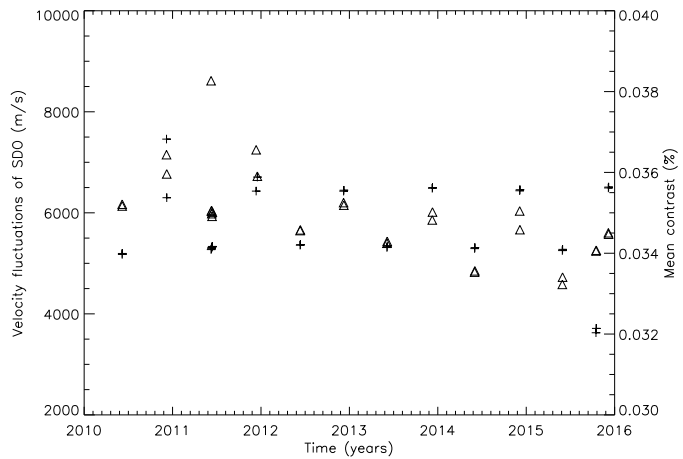}
 \caption[]{ Time evolution of the intensity contrasts (triangles)
and of fluctuations of the velocity module of the SDO satellite
(crosses).} 
\label{Velcont}
\end{figure}

 Observations over the entire disk of the longitudinal
magnetic field and Doppler velocities are provided by SDO/HMI.

Figure~\ref{Bmoy} shows no particular variation trend of the
magnitude of the magnetic field during the cycle (which corroborates
results of Figure~\ref{bmagcyc} at varying $B_0$). The larger value
of $|BLOS|$ on December 4, 2014, is due to a small residual active
region in the upper right corner of the field. Figure~\ref{Bmoy} also displays 
the magnitude of magnetic fields in regions where
$|BLOS| < 100 G$. We observe a better stability in very quiet
regions of the Sun.

 Continuum intensity contrasts of Figure~\ref{Bmoy} are derived
from the average of the contrast of each image in the 6-hour
sequence. Contrast was corrected from pixel size fluctuations
(3\%) due mainly to the varying apparent radius.
 However, the continuum intensity is one of the outputs of a fit to the
measurements at the six tuning positions of HMI during the line
scan. When orbital or solar velocities are large this can introduce
artifacts in the continuum intensity. As shown in
Figure~\ref{Velcont}, the constrast intensity is well correlated with
velocity fluctuations of the SDO satellite. The intensity contrast
is clearly modulated by SDO proper motions giving a lower variation
than those plotted in figure ~\ref{Bmoy}. In conclusion, the
intensity contrast does not reveal any variation in time,
corroborating preliminary conclusions of section 2 and previous
results found by \citet{MHS2007} and Muller (private communication).

 \subsection{Doppler temporal evolution}

\begin{figure}
\centering
\includegraphics[width=9cm]{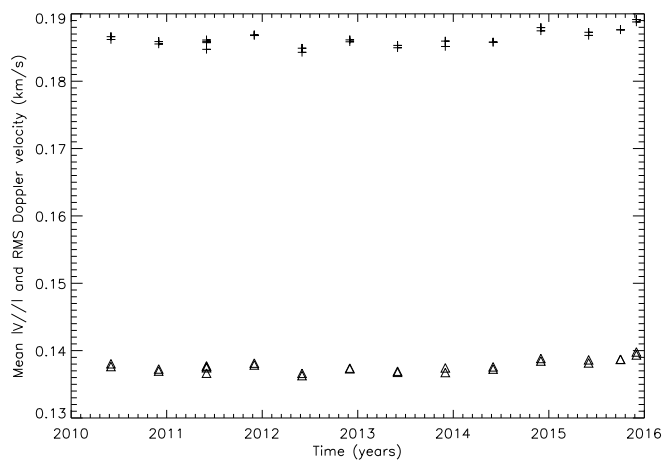}
 \caption[]{Plus signs: mean value of VLOS magnitude (km/s); triangles: RMS VLOS
 (km/s).
}
\label{Vmdop}
\end{figure}

The Doppler velocity was obtained at disk center for 1 hour over
a 216.4 Mm~ $\times$~ 216.4 Mm. FOV  (Table 1) at zero $B_0$,
in order to avoid the annual modulation described above for BLOS
(section 2).  The HMI Doppler signal is formed at a height of approximately 
100 km (\citep{FCS2011}). 

The line of sight Doppler velocity VLOS was first
corrected from 5 min oscillations in the $k$ and $\omega$ diagram
using the classical filter $ \omega < V_{\rm cut-off} \times k$
where $V_{\rm cut-off}=6\,\mathrm{km~s}^{-1}$ in order to isolate the
convective component.

VLOS was then corrected in latitude and longitude from the solar
rotation using a differential rotation law of the form (in degrees
per day) :

\begin{equation}
\Omega =  14.41 - 2.83 * \sin^2 \theta 
\end{equation}

The latitude $\theta$, proportional to y, allows computation of the angular
velocity as a function of y. Hence, the projection of the rotation
speed on the LOS is a function of x and y which was subtracted from
VLOS. Thus VLOS has a zero average value over the FOV.

Figure~\ref{Vmdop} shows that we do not detect any VLOS variation
along the solar cycle.

\section{Discussion and conclusions}

 The solar cycle was originally found through the variations in the number of sunspots,
which are easily observable. Besides, we know that the quiet
Sun is far from devoid of magnetic field. Changes in the dynamics of
the quiet Sun could have a large effect on its overall properties, such
as the photospheric temperature or magnetic field generation (flux
tubes). In many cases, the existence of variations with the solar
cycle of various quiet Sun parameters, has yet to be demonstrated.
  Understanding the mechanism that diffuses magnetic fields  over the solar surface is
still an important challenge, because the network contribution to
the global solar magnetism is known to be comparable to the flux of
active regions at maximum activity. As the global  magnetic field of
the Sun has an activity cycle, one expects to observe some
variations in the dynamical properties of the flows in the
photosphere. The comparison with different phases of the solar cycle
studied by \citet{PJP1995} reveals a high stability of the solar
granulation while mesogranulation is dependent.
 As some variations of the supergranulation \citep{RS1989} and granulation (Muller,
private communication) are observed with latitude, our measurements
at disk center reveal the stability of the flow properties between
meso- and supergranular scales along the solar cycle. In our study,
we took care to select very quiet regions without any residual
activity. In such regions, the detected magnetic field is probably
generated by the local dynamo process to form the quiet network.
\citet{UMTV2016} describe a link between the magnetic bright point
(MBP) activity close to the equator and the global magnetic cycle.
They indicate that a significant fraction of MBPs originate from
the sunspot activity belt. Thus the magnetic network behavior appears to be 
related to solar activity but its variation is essentially fed by
the decay of active regions \citep{TCB2014}. Our analysis seems to
indicate that the magnetic field produced locally at disk center
\citep{GBOKD14}, independently from the large scale dynamo, forms
what we call the very quiet Sun, and remains unchanged along the solar
cycle. Thus the irradiance contribution of the very quiet network is also likely 
be constant in time. The latitudinal variation
of the network properties \citep{ITJ2010} and related irradiance,
could be essentially due to the magnetic field coming from active
regions dispersed by differential rotation, meridian and
supergranular flows \citep{DGG2015}.

 The present work has been concentrated on large scales as meso$-$ and
super$-$ granulation; further analysis will be extended to the
solar granulation dynamics along the cycle. Indeed, the observed 
parameters of the quiet Sun (intensity, velocity, and field strength) 
are constrained by the spatial resolution of HMI. This does not a priori 
exclude that changes could be occurring at smaller spatial scales than 
meso- and supergranulation.

\begin{acknowledgements}
 We thank the anonymous referee for his/her careful reading of
our manuscript and his/her many insightful comments and
suggestions. The data used here are courtesy of NASA/SDO and the
HMI Science Team, which we thank for their support. This work was
granted access to the HPC resources of CALMIP under the allocation
2011-[P1115]. Particular thanks to B. Stein for providing his
simulations, to Wilcox Solar Observatory for polar field and mean
field data and to Brussels Royal Observatory for the sunspot index.

\end{acknowledgements}

\bibliographystyle{aa}
\bibliography{biblio}

\begin{thebibliography}{28}
\expandafter\ifx\csname natexlab\endcsname\relax\def\natexlab#1{#1}\fi

\bibitem[{{Dasi-Espuig} {et~al.}(2016){Dasi-Espuig}, {Jiang}, {Krivova},
  {Solanki}, {Unruh}, \& {Yeo}}]{DEJK2016}
{Dasi-Espuig}, M., {Jiang}, J., {Krivova}, N.~A., {et~al.} 2016, \aap, 590, A63

\bibitem[{{DeRosa} \& {Toomre}(2004)}]{DET04}
{DeRosa}, M.~L. \& {Toomre}, J. 2004, \apj, 616, 1242

\bibitem[{{Fleck} {et~al.}(2011){Fleck}, {Couvidat}, \& {Straus}}]{FCS2011}
{Fleck}, B., {Couvidat}, S., \& {Straus}, T. 2011, \solphys, 271, 27

\bibitem[{{Georgobiani} {et~al.}(2007){Georgobiani}, {Zhao}, {Kosovichev},
  {Benson}, {Stein}, \& {Nordlund}}]{GZKB2007}
{Georgobiani}, D., {Zhao}, J., {Kosovichev}, A.~G., {et~al.} 2007, \apj, 657,
  1157

\bibitem[{{Giannattasio} {et~al.}(2014){Giannattasio}, {Stangalini},
  {Berrilli}, {Del Moro}, \& {Bellot Rubio}}]{GSBDB2014}
{Giannattasio}, F., {Stangalini}, M., {Berrilli}, F., {Del Moro}, D., \&
  {Bellot Rubio}, L. 2014, \apj, 788, 137

\bibitem[{{Gizon} \& {Duvall}(2004)}]{GD2004}
{Gizon}, L. \& {Duvall}, T.~L. 2004, in IAU Symposium, Vol. 223,
  Multi-Wavelength Investigations of Solar Activity, ed. A.~V. {Stepanov},
  E.~E. {Benevolenskaya}, \& A.~G. {Kosovichev}, 41--44

\bibitem[{{Go{\v s}i{\'c}} {et~al.}(2014){Go{\v s}i{\'c}}, {Bellot Rubio},
  {Orozco Su{\'a}rez}, {Katsukawa}, \& {del Toro Iniesta}}]{GBOKD14}
{Go{\v s}i{\'c}}, M., {Bellot Rubio}, L.~R., {Orozco Su{\'a}rez}, D.,
  {Katsukawa}, Y., \& {del Toro Iniesta}, J.~C. 2014, \apj, 797, 49

\bibitem[{{Ishikawa} {et~al.}(2010){Ishikawa}, {Tsuneta}, \& {Jur{\v
  c}{\'a}k}}]{ITJ2010}
{Ishikawa}, R., {Tsuneta}, S., \& {Jur{\v c}{\'a}k}, J. 2010, \apj, 713, 1310

\bibitem[{{Meunier} {et~al.}(2008){Meunier}, {Roudier}, \&
  {Rieutord}}]{MRR2008}
{Meunier}, N., {Roudier}, T., \& {Rieutord}, M. 2008, \aap, 488, 1109

\bibitem[{{Meunier} {et~al.}(2007){Meunier}, {Roudier}, \& {Tkaczuk}}]{MRT2007}
{Meunier}, N., {Roudier}, T., \& {Tkaczuk}, R. 2007, \aap, 466, 1123

\bibitem[{{Meunier} \& {Zhao}(2009)}]{MZ2009}
{Meunier}, N. \& {Zhao}, J. 2009, \ssr, 144, 127

\bibitem[{{Muller} {et~al.}(2007){Muller}, {Hanslmeier}, \&
  {Salda{\~n}a-Mu{\~n}oz}}]{MHS2007}
{Muller}, R., {Hanslmeier}, A., \& {Salda{\~n}a-Mu{\~n}oz}, M. 2007, \aap, 475,
  717

\bibitem[{{Muller} {et~al.}(2006){Muller}, {Salda{\~n}a-Mu{\~n}oz}, \&
  {Hanslmeier}}]{MSH2006}
{Muller}, R., {Salda{\~n}a-Mu{\~n}oz}, M., \& {Hanslmeier}, A. 2006, Advances
  in Space Research, 38, 891

\bibitem[{November \& Simon(1988)}]{NS88}
November, L.~J. \& Simon, G.~W. 1988, ApJ, 333, 427

\bibitem[{{Palle} {et~al.}(1995){Palle}, {Jimenez}, {Perez Hernandez},
  {Regulo}, {Roca Cortes}, \& {Sanchez}}]{PJP1995}
{Palle}, P.~L., {Jimenez}, A., {Perez Hernandez}, F., {et~al.} 1995, \apj, 441,
  952

\bibitem[{{Rieutord} \& {Rincon}(2010)}]{RR10}
{Rieutord}, M. \& {Rincon}, F. 2010, Living Reviews in Solar Physics, 7, 2

\bibitem[{{Rimmele} \& {Schroeter}(1989)}]{RS1989}
{Rimmele}, T. \& {Schroeter}, E.~H. 1989, \aap, 221, 137

\bibitem[{{Roudier} \& {Reardon}(1998)}]{RR1998}
{Roudier}, T. \& {Reardon}, K. 1998, in Astronomical Society of the Pacific
  Conference Series, Vol. 140, Synoptic Solar Physics, ed. K.~S.
  {Balasubramaniam}, J.~{Harvey}, \& D.~{Rabin}, 455

\bibitem[{Roudier {et~al.}(1999)Roudier, Rieutord, Malherbe, \&
  Vigneau}]{RRMV99}
Roudier, T., Rieutord, M., Malherbe, J., \& Vigneau, J. 1999, \aap, 349, 301

\bibitem[{{Scherrer} {et~al.}(2012){Scherrer}, {Schou}, {Bush}, {Kosovichev},
  {Bogart}, {Hoeksema}, {Liu}, {Duvall}, {Zhao}, {Title}, {Schrijver},
  {Tarbell}, \& {Tomczyk}}]{Scherrer2012}
{Scherrer}, P.~H., {Schou}, J., {Bush}, R.~I., {et~al.} 2012, \solphys, 275,
  207

\bibitem[{{Schou} {et~al.}(2012){Schou}, {Scherrer}, {Bush}, {Wachter},
  {Couvidat}, {Rabello-Soares}, {Bogart}, {Hoeksema}, {Liu}, {Duvall}, {Akin},
  {Allard}, {Miles}, {Rairden}, {Shine}, {Tarbell}, {Title}, {Wolfson},
  {Elmore}, {Norton}, \& {Tomczyk}}]{Schou2012}
{Schou}, J., {Scherrer}, P.~H., {Bush}, R.~I., {et~al.} 2012, \solphys, 275,
  229

\bibitem[{{Stein}(2012)}]{Stein2012}
{Stein}, R.~F. 2012, Living Reviews in Solar Physics, 9

\bibitem[{{Stein} {et~al.}(2009){Stein}, {Nordlund}, {Georgoviani}, {Benson},
  \& {Schaffenberger}}]{Stein2009}
{Stein}, R.~F., {Nordlund}, {\AA}., {Georgoviani}, D., {Benson}, D., \&
  {Schaffenberger}, W. 2009, in Astronomical Society of the Pacific Conference
  Series, Vol. 416, Solar-Stellar Dynamos as Revealed by Helio- and
  Asteroseismology: GONG 2008/SOHO 21, ed. M.~{Dikpati}, T.~{Arentoft},
  I.~{Gonz{\'a}lez Hern{\'a}ndez}, C.~{Lindsey}, \& F.~{Hill}, 421

\bibitem[{{Thibault} {et~al.}(2014){Thibault}, {Charbonneau}, \&
  {B{\'e}land}}]{TCB2014}
{Thibault}, K., {Charbonneau}, P., \& {B{\'e}land}, M. 2014, \apj, 796, 19

\bibitem[{{Title} {et~al.}(1989){Title}, {Tarbell}, {Topka}, {Ferguson},
  {Shine}, \& {SOUP Team}}]{TTTFS89}
{Title}, A.~M., {Tarbell}, T.~D., {Topka}, K.~P., {et~al.} 1989, ApJ, 336, 475

\bibitem[{{Utz} {et~al.}(2016){Utz}, {Muller}, {Thonhofer}, {Veronig},
  {Hanslmeier}, {Bodn{\'a}rov{\'a}}, {B{\'a}rta}, \& {del Toro
  Iniesta}}]{UMTV2016}
{Utz}, D., {Muller}, R., {Thonhofer}, S., {et~al.} 2016, \aap, 585, A39

\bibitem[{{van Driel-Gesztelyi} \& {Green}(2015)}]{DGG2015}
{van Driel-Gesztelyi}, L. \& {Green}, L.~M. 2015, Living Reviews in Solar
  Physics, 12

\bibitem[{{Yeo} {et~al.}(2014){Yeo}, {Feller}, {Solanki}, {Couvidat},
  {Danilovic}, \& {Krivova}}]{YEO2014}
{Yeo}, K.~L., {Feller}, A., {Solanki}, S.~K., {et~al.} 2014, \aap, 561, A22

\end{thebibliography}

\end{document}